\documentclass[a4paper]{jpconf}
\usepackage{graphicx}
\begin{document}
\title{Prospects for heavy flavour measurements with the ALICE inner tracker upgrade}

\author{Cristina Terrevoli (for the ALICE Collaboration)}

\address{Univerity and INFN Cagliari, I-09042 Monserrato (CA)
ITALY }

\ead{cristina.terrevoli@cern.ch}

\begin{abstract}

ALICE is the general purpose heavy-ion detector at the CERN LHC. Its goal is to investigate the properties
of the strongly interacting matter under the extreme conditions of density and temperature reached 
in Pb--Pb collisions, with the aim to characterize the Quark-Gluon Plasma (QGP). 
In this scenario, the upgrade of the ALICE inner tracker targets physics topics in which ALICE
can bring a unique contribution to the QGP characterization via the heavy flavour probes.
We present an overview of the inner tracker upgrade and the expected
physics performance for heavy flavour measurements.

\end{abstract}

\section{Introduction}

Among the probes useful to investigate the properties of the QGP, heavy quarks play a special role because
they are produced in the very initial stage of the collisions and they carry information about the
properties of the traversed medium. An accurate measurement of heavy flavour
provides information on fundamental properties of the medium, such as the transport coefficient, the
thermalization and hadronization mechanisms.
Interesting results have been obtained in the first three years of data-taking at the LHC, but
there are still open questions, which would require measurements beyond the present
capabilities of the ALICE apparatus. Among them, the most interesting are the measurements of
the nuclear modification factor $(R_ {\rm{AA}})$ and anisotropy of the azimuthal distributions of charm
and beauty mesons down to transverse momentum below 1 GeV/$c$. Another completely
unexplored field is the measurement of the production of heavy flavour baryons, like the
$\mathrm{\Lambda_c}$, that can bring insight on the heavy quark hadronization mechanism
in the presence of a partonic medium.
At present such measurements are limited by the resolution of the Inner Tracking System (ITS), which,
in particular, is not sufficient to measure in Pb--Pb collisions the production of $\mathrm{\Lambda_c}$
baryons, that have a mean proper decay length of only 60 $\mu$m. Another limitation of the present
ALICE central barrel detectors to the measurement of heavy flavours at low momentum is the
maximum achievable readout rate, which prevents the full exploitation of the luminosity
expected to be delivered by the LHC after the Long Shutdown 2.
An upgrade of the inner tracking system based on today’s frontier technologies 
will improve the current performance in the
pointing and momentum resolution, providing in addition a high tracking efficiency down to
very low transverse momentum. Moreover, a faster readout, for all the central barrel detectors,
will allow for a data collection rate, in Pb--Pb collisions, a factor 100 larger than at present, and
this will contribute to enhance the ALICE physics performance very significantly.
\section{ITS Upgrade concept}
The current ITS has been designed to have an excellent capability 
to separate primary and secondary vertices of 
heavy flavour hadrons. It is composed of six layers of silicon detectors
(pixels, drifts, and strips) with the innermost layer at a radius of 3.9 cm.
The features of the ITS Upgrade \cite{CDR} as compared with the present ITS
are the following: (i) the innermost layer closer to the beam line at 2.2 cm, 
(ii) reduced 
material budget from the current 1.1\% per pixel layer, to $\sim$0.3\%, 
(iii) smaller size pixel cells (from $50\times425 \mu$m to (20-30)$\times$(20-30)$\mu$m), (iv) a seventh detector layer. 
This will allow for a significant improvement of the tracking performance and the momentum resolution.
In particular, the impact parameter resolution 
 will be improved by a factor of three in the transverse direction and a factor of six in the longitudinal direction.
 Furthermore a continuous readout of Pb--Pb interactions at $\sim$50 kHz will permit to exploit
the upgrade LHC luminosity after 2018.
The upgrade is targeted for the second long shutdown in 2017-2018.

\section{Heavy Flavour results with
the current Inner Tracking System} 

\begin{figure}[t!]
\begin{minipage}{15pc}
\includegraphics[width=15pc]{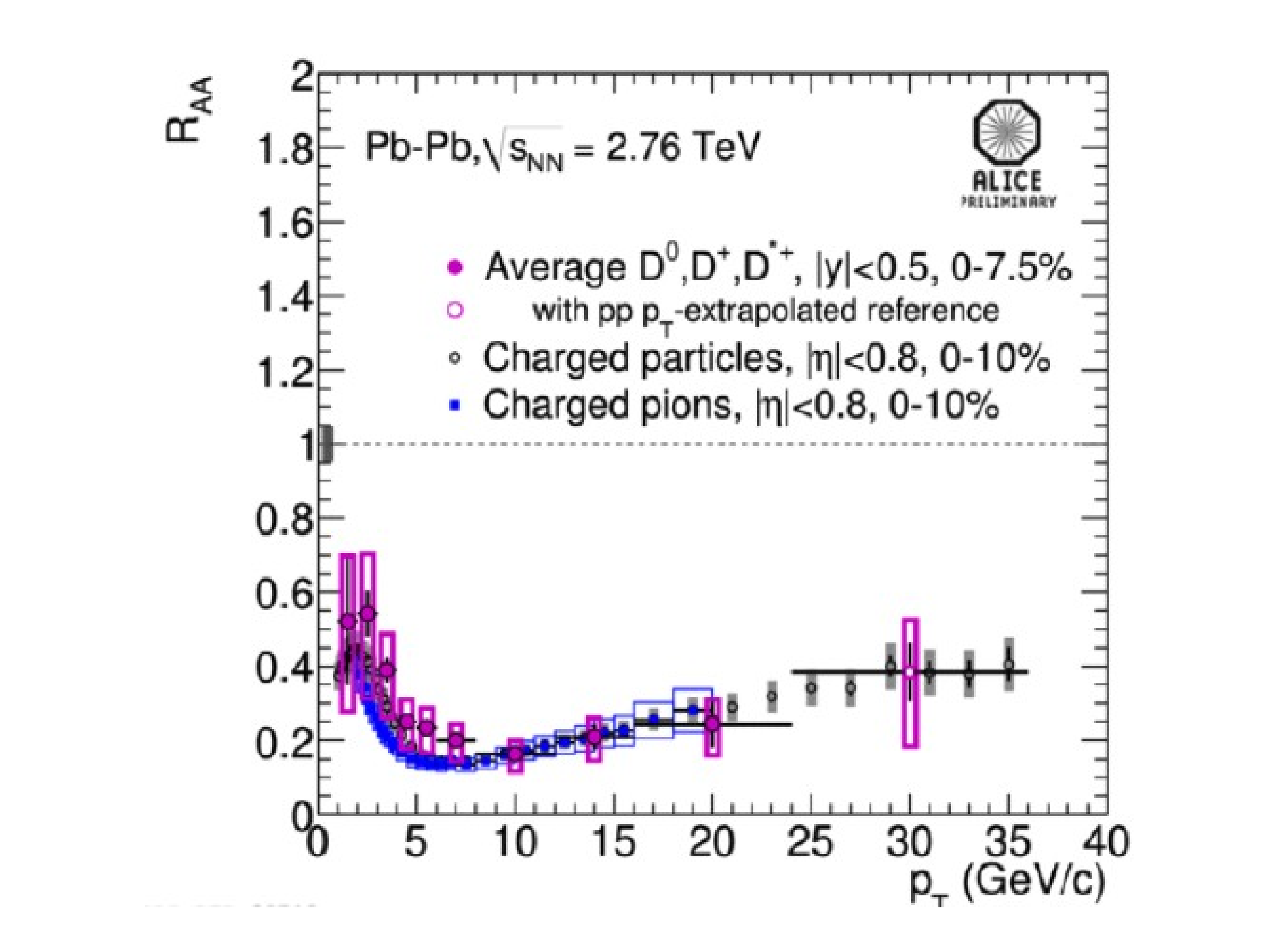}
\caption{\label{raadmes} $R_{\rm{AA}}$ of D mesons compared with the $R_{\rm{AA}}$ of charged pions in Pb--Pb collisions at $\sqrt{s_{NN}}$= 2.76 TeV.}
\end{minipage}\hspace{4pc}%
\begin{minipage}{15pc}
\includegraphics[width=15pc]{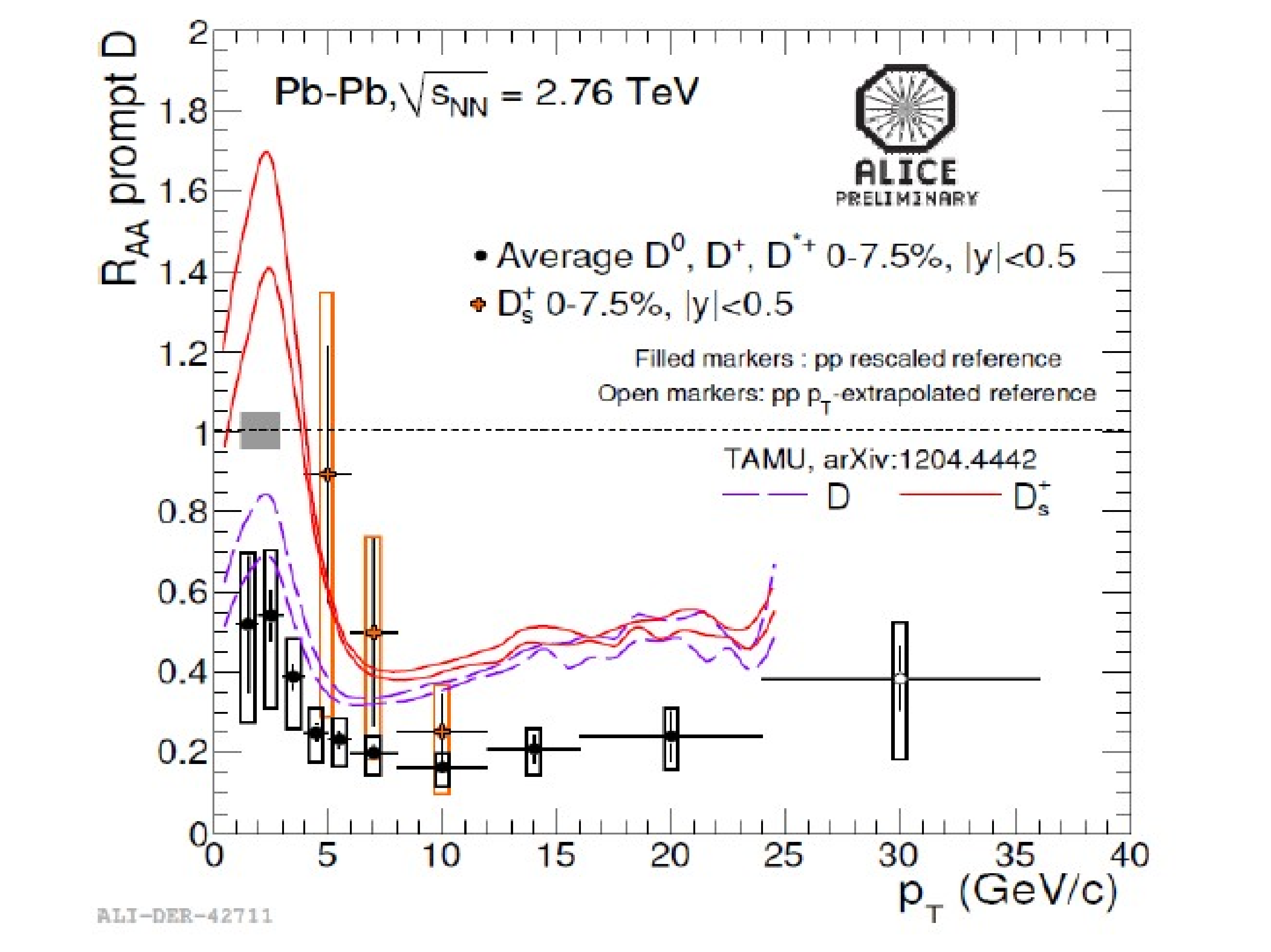}
\caption{\label{dores}$\mathrm{D_{s}}$ $R_{\rm{AA}}$ compared with the theorethical models and D mesons $R_{\rm{AA}}$. }
\end{minipage} 
\end{figure}

The energy-loss mechanism for
different parton species is among the observables useful to investigate the properties of the QGP matter.
Theoretical models predict a dependence on the colour charge and mass
of partons propagating through the medium. In particular, a larger energy loss is expected for gluons,
together with a suppression of gluon radiation at small angles for partons with larger mass (dead cone effect). 
This dependences can be investigated using the nuclear modification factor $R_{\rm AA }$, 
the ratio of the $p_{\rm T}$ distribution in Pb--Pb and the $p_{\rm T}$ 
distribution in pp, scaled by the number of binary nucleon-nucleon collisions. 
The expected pattern is  $R_{\rm AA}^\pi < R_{\rm AA}^D < R_{\rm AA}^B$ \cite{enloss}.
The $R_{\rm AA}$ of D mesons ($\mathrm{D^0}$, $\mathrm{D^+}$, $\mathrm{D^{*+}}$)
was measured in central Pb--Pb collisions at $\sqrt{s_{NN}}$ =2.76 TeV and compared with that of charged pions \cite{DAlice}, as shown
in Fig.~\ref{raadmes}.
The data suggest that the suppression of D mesons might be smaller than for pions in the low momentum region, 
but the large systematic uncertainties prevent a firm conclusion.
The measurement of the B mesons $R_{\rm{AA}}$ is not accessible with the current setup of the ALICE detector. 
The upgrade of the ITS will open the possibility to study B mesons down to zero $p_{\rm T}$.
 Moreover the improved tracking resolution
and efficiency, together with the higher statistics, 
will improve the $p_{\rm T}$ coverage and the accuracy of D meson measurements.
Another open question concerning the heavy-flavour interaction with the QGP medium, involves the 
hadronization of heavy quarks in the medium, which can be studied by measuring the 
baryon/meson ratio for charm and beauty ($\mathrm{\Lambda_c}/\mathrm{D}$ and $\mathrm{\Lambda_b}/\mathrm{B}$):
coalescence models predict an increase of baryon-to-meson ratio \cite{coal}.
With the current set-up of ALICE, the analysis of $\mathrm{\Lambda_c}$ and $\mathrm{\Lambda_b}$ is not accessible in Pb--Pb
due to the limited precision and statistics.
Another important test for coalescence models is that, if coalescence 
contributes to the charm hadronization, the $\mathrm{D_s}$ is expected to be enhanced with respect to the other 
D mesons at low $p_{\rm T}$.
There is an hint of this enhancement \cite{DsGian}(see Fig.~\ref{dores}), but it is necessary
to improve tracking precision, statistics and to extend the measurement to low $p_{\rm T}$ to have a conclusive 
answer.
The measurement of the elliptic flow $v_2$ is sensitive to the thermalitazion of charm and beauty in the QGP.
Models predict a large D meson $v_2$ at low momentum and a mass dependence of $v_2$ ( $v_2(\mathrm{B}) <  v_2(\mathrm{D})$)
\cite{v2,bampsv2}. 
The ALICE measurement of D meson $v_2$ down to 2 GeV/$c$ shows a positive value of $v_2$ 
at intermediate $p_{\rm T}$ in semi-central collisions,  but with a limited precision. The measurement is not accessible for B mesons.
The new ITS will allow for a precise measurement of the D meson $v_2$ down to zero $p_{\rm T}$ and will open the possibility 
for the measurement of B meson $v_2$.
\section{\label{ITSUpgrade}ITS Upgrade physics performance}
\begin{figure}[h]
\begin{minipage}{14pc}
\includegraphics[width=14pc]{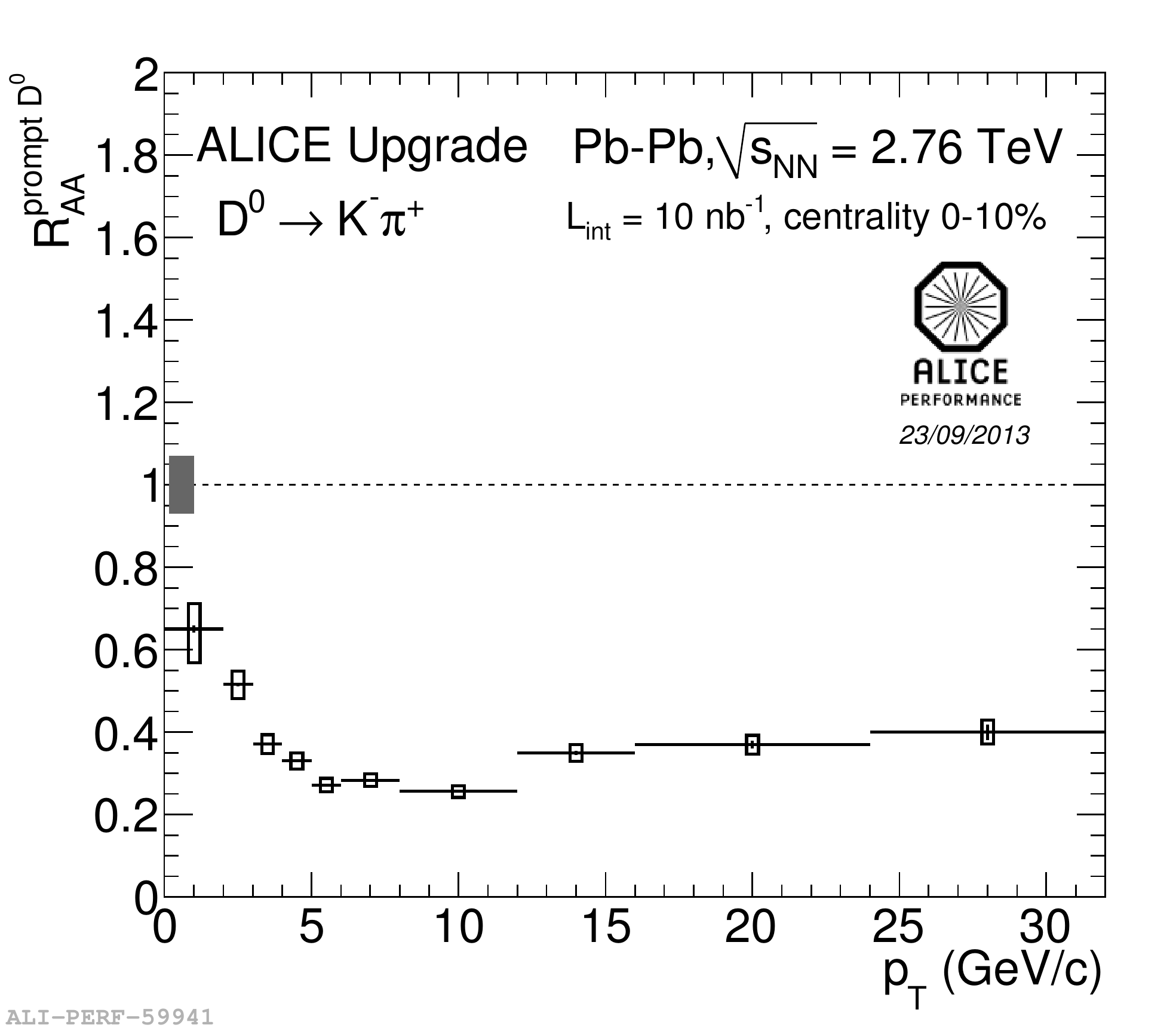}
\caption{\label{d0} $R_{AA}$ of $\mathrm{D^0}$ in Pb--Pb at 5.5 TeV with the ALICE upgrade.}
\end{minipage}\hspace{3pc}
\begin{minipage}{15pc}
\includegraphics[width=15pc]{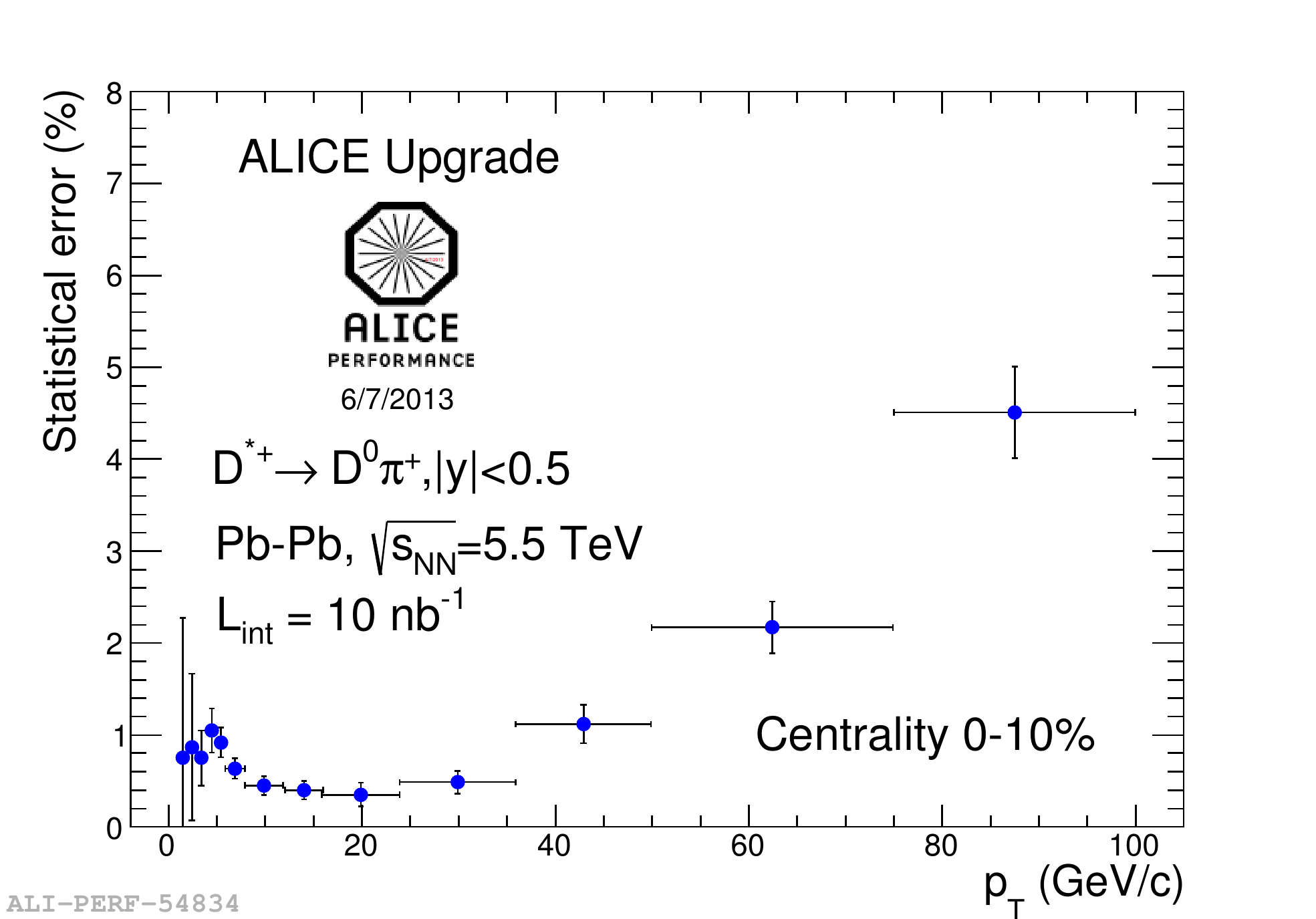}
\caption{\label{ds} Statistical uncertainties of the $\mathrm{D^*}$ measurement in Pb--Pb 5.5 TeV with the ALICE upgrade.}
\end{minipage} 
\end{figure}

\begin{figure}[h]
\includegraphics[width=0.9\linewidth]{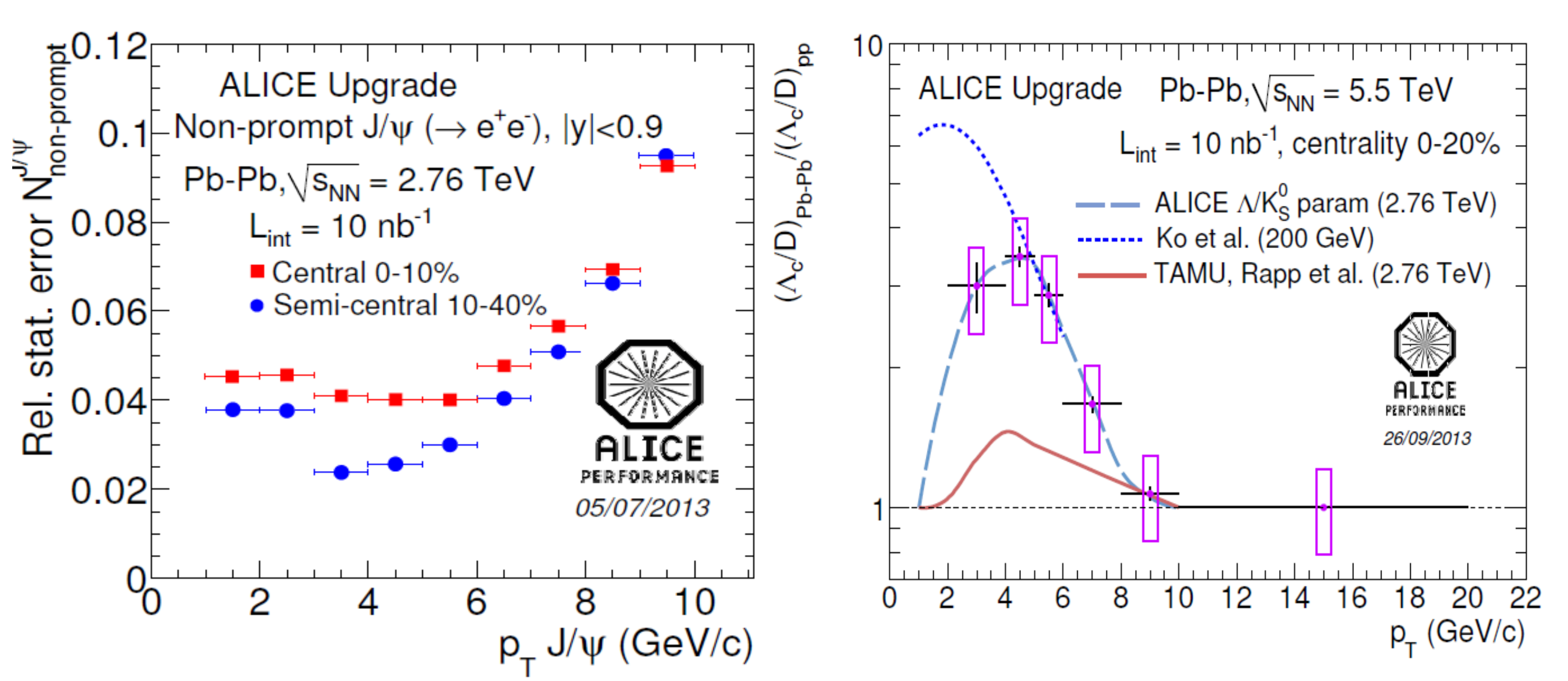}
\caption{\label{jpsi}(Left) Non-prompt $\mathrm{J}$/$\mathrm{\psi}$ yield: statistical uncertainty, for central and semi-central collisions, with the ALICE upgrade.
(Right) $\mathrm{\Lambda_c}$/$\mathrm{D^0}$ ratio in Pb--Pb vs pp collisions, measured with the ALICE upgrade.}
\end{figure}

Several simulation studies have been carried out to quantify the improved performance of the upgraded ITS
and the higher statistics achievable thanks to the continuous read-out (up to 10~nb$^{-1}$ of integrated luminosity).     
The D meson measurement will be extended at low and high $p_{\rm T}$, with improved statistical precision.
 In Fig.~\ref{d0} the expected $R_{\rm{AA}}$ of $\mathrm{D^0}$ mesons is shown: the upgraded ITS allows to reach down to zero
$p_{\rm T}$.   
In Fig.~\ref{ds} the statistical uncertainty of the $\mathrm{D^{*+}}$ measurement is shown. The analysis 
can be carried out in a wider range (currently 3-36 GeV/$c$) with high significance (20-100),
taking advantage of the higher precision and statistics.
A direct measurement of beauty will be possible 
via non-prompt $\mathrm{D^0}$ and $\mathrm{J}/\mathrm{\psi}$, as shown in Fig.~\ref{jpsi} (left),
where the statistical uncertainty for non-prompt $\mathrm{J}/\mathrm{\psi}$ is presented. This will allow for the measurement
of the $v_2$ of B mesons for the first time.
The $\mathrm{\Lambda_c}$ baryon will be accessible for $p_{\rm T} >$2 GeV/$c$ in Pb--Pb thanks to the improved resolution, with a higher precision. In 
 Fig.~\ref{jpsi} (right) the double ratio $\mathrm{\Lambda_c}$/$\mathrm{D}$ in Pb--Pb over pp is shown together with two theoretical predictions.
Studies on the $\mathrm{\Lambda_b}$ reconstruction are also ongoing: the baryon will be accessible for the first time via its decay in $\mathrm{\Lambda_c}$. 

\begin{figure}[t!]
\includegraphics[width=0.9\linewidth]{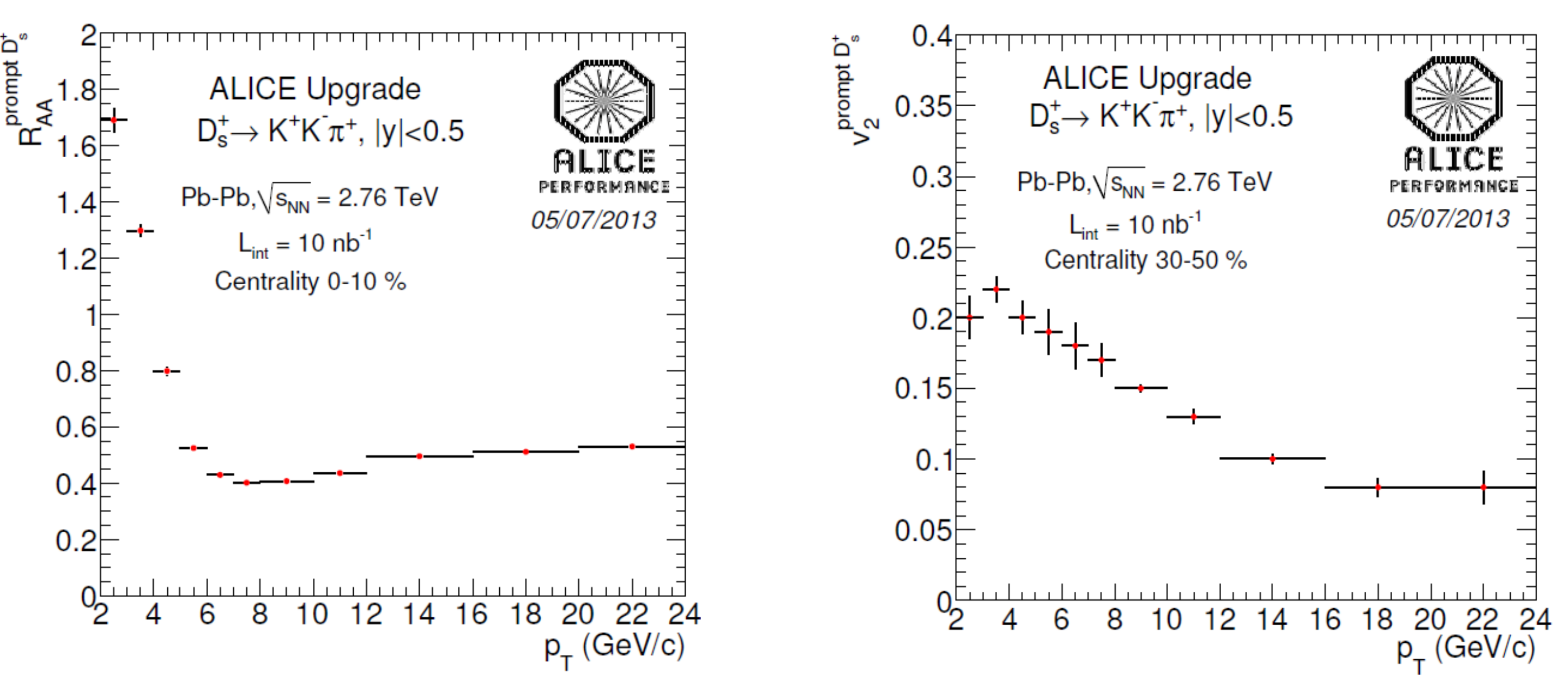}
\caption{\label{dsup}(Left) Statistical uncertainty on $R_{\rm{AA}}$ for $\mathrm{D_s}$ with the ALICE upgrade. (Right) $v_2$ for $\mathrm{D_s}$ with the ALICE upgrade.}
\end{figure}

Also the $\mathrm{D_s}$ analysis will benefit for the much larger available statistics, with expected precise measurements of $R_{\rm_{AA}}$ and $v_2$ down to low
$p_{\rm T}$ (see Fig.~\ref{dsup}).

\section{Conclusions}

ALICE has a strong upgrade physics programme for precision QGP studies, in particular for a deeper 
understanding of energy loss mechanisms, azimuthal anisotropy
and in-medium hadronization, where heavy flavour measurements play a central role.
The main requirements for the upgrade are enhanced rate capabilities and a new Inner Tracking System.
These features will provide a strong increase of the statistical precision in the measurements of yields and spectra of heavy flavour mesons and baryons.


\section*{References}
\begin{thebibliography}{5}
\bibitem{CDR}  ALICE Collaboration, 2012 ITS Upgrade Conceptual Design Report {\it CERN-LHCC-2012-013}
\bibitem{enloss} Dokshitzer Y L and Kharzeev D, et al 2001 {\it Phys. Lett.} B {\bf 519} 199.
\bibitem{DAlice} Abelev B et al 2012 (ALICE Collaboration) arXiv:1203.2160.
\bibitem{DsGian} Innocenti G et al (ALICE Collaboration) 2012  {\it Quark Matter 2012} http://qm2012.bnl.gov
\bibitem{coal}Lee S H, Ohnishi K, Yasui S, Yoo I and Ko C, 2008 {\it Phys. Rev. Lett.} {\bf 100} 222301 
\bibitem{v2} Aichelin J, Gossiaux P and Gousset T 2012 arXiv:1201.4192
\bibitem{bampsv2} Uphoff J, Fochler O, Xu Z and Greiner C 2012 arXiv:1205.4945

\end{thebibliography}
\end{document}